\documentclass[aps,pra,twocolumn,superscriptaddress,10pt,floatfix]{revtex4-2}
\usepackage{multirow}
\usepackage{xcolor}
\usepackage{makecell}
\usepackage{graphicx}
\usepackage{dcolumn}
\usepackage{bm}
\usepackage{natbib}
\usepackage{url}
\usepackage{hyperref}
\usepackage[utf8]{inputenc}
\usepackage{amsmath}
\usepackage{hyperref}
\usepackage{nameref}   
\usepackage{cleveref}

\begin{document}

\preprint{APS/Proton_on_Coronene}

\title{Multiple ionization, fragmentation and dehydrogenation of coronene in collisions with swift protons}

\author{Shashank Singh}
\affiliation{Department of Nuclear and Atomic Physics, Tata Institute of Fundamental Research, Dr. Homi Bhabha Road, Colaba, Mumbai 400005, India}

\author{Sanjeev Kumar Maurya}
\affiliation{Department of Nuclear and Atomic Physics, Tata Institute of Fundamental Research, Dr. Homi Bhabha Road, Colaba, Mumbai 400005, India}

\author{Shikha Chandra}
\affiliation{Department of Nuclear and Atomic Physics, Tata Institute of Fundamental Research, Dr. Homi Bhabha Road, Colaba, Mumbai 400005, India}

\author{Debasmita Chakraborty}
\affiliation{Department of Nuclear and Atomic Physics, Tata Institute of Fundamental Research, Dr. Homi Bhabha Road, Colaba, Mumbai 400005, India}

\author{Laszlo Gulyás}
\affiliation{Institute of Nuclear Research of the Hungarian Academy of Sciences (ATOMKI), H-4001 Debrecen, Hungary}

\author{Lokesh C Tribedi}
\thanks{email:\ lokesh@tifr.res.in, \ ltribedi@gmail.com (corresponding author)}
\affiliation{Department of Nuclear and Atomic Physics, Tata Institute of Fundamental Research, Dr. Homi Bhabha Road, Colaba, Mumbai 400005, India}

\affiliation{University of Petroleum and Energy studies(UPES), Bidholi, Dehradun 248007, Uttarakhand, India}


\date{\today}

\begin{abstract}
 
{Coronene molecules have been bombarded with protons of energy ranging from 100 to 300 keV. The time-of-flight mass spectra have been recorded using a two-stage Wiley-McLaren-type spectrometer. A significant enhancement in the yields of doubly and triply ionized recoil-ions is observed compared to the singly ionized ones. The single, double and triple ionization cross-sections are also calculated theoretically using the continuum distorted wave–eikonal initial state (CDW-EIS)  and are compared with the experimental results. The experimental ratios of yields of double-to-single charged and triple-to-single charged recoil-ions are found to be much higher compared to those for the gaseous atoms. Evaporation peaks corresponding to the loss of several neutral $C_2H_2$ molecules are observed for singly, doubly and triply charged coronene recoil ions. Multi-fragmentation peaks corresponding to smaller masses of carbohydrates $C_nH_x^+$ (n = 3–7), appear in the spectra due to higher energy transfer from the projectile to the molecule. The yields of evaporation and fragment products exhibit a pronounced dependence on projectile energy, with a significant decrease observed at higher energies. Dehydrogenetaion i.e. loss of H-atoms or H$_2$ molecules are also investigated from the measured spectra. It is observed that hydrogen molecule losses are preferred over H-loss in the cation and dication coronene peak structures, with up to three molecules being lost. This observation is in line with some of the predictions and may provide important inputs towards the astrochemistry regarding the observed abundance of H$_2$ in the inter stellar medium.}
\end{abstract}

\maketitle


\section{Introduction}
\par 
Polycyclic aromatic hydrocarbon (PAH) molecules are widely observed in the universe through their distinct spectroscopic signatures. In the optical region (0.4–1.3 $\mu$m), they are identified via diffuse interstellar bands (DIBs), which are detected along the lines of sight toward stars \cite{joblin1990detection, herbig1995diffuse, salama1996assessment}. In the infrared region, PAHs are recognized through aromatic infrared bands (AIBs), which arise from vibrationally excited PAHs emitting in response to ultraviolet-visible and near-infrared radiation \cite{allamandola1985polycyclic, allamandola1989interstellar, joblin1990detection}. In addition to their role in DIBs and AIBs, PAHs are also considered potential carriers of the extended red emission (ERE), observed in the 540–900 nm range \cite{witt2008extended, berne2008extended, rhee2007charged}. Quantum chemical calculations by Rhee et al. \cite{rhee2007charged} suggest that PAH dimers (\([PAH_2]^+\)) could be responsible for this emission. Furthermore, Moreels et al. \cite{moreels1994detection} proposed that PAHs are the origin of the unidentified fluorescence bands observed in the 280–480 nm range by the three-channel spectrometer (TKS) onboard the Vega 2 spacecraft in the coma of comet Halley. Allamandola et al., Leger and Puget \cite{allamandola1985polycyclic, leger1984identification} proposed models to identify the presence of PAHs in the interstellar medium. Later supported by Allamandola et al. \cite{allamandola1999modeling}, who modeled the infrared band spectrum of various interstellar objects using laboratory data from both neutral and ionized PAHs. Today, PAHs are known to exist not only in the Milky Way but also in extragalactic sources, the local interstellar medium, outer solar system bodies, meteorites, interplanetary dust particles, reflection nebulae, planetary nebulae and other interstellar and intergalactic environments.

\par The PAH molecules consist of two or more aromatic rings and are stabilized by the delocalization of \(\pi\) electrons. Their high ionization potentials and dissociation energies (of the order of tens of eV \cite{holm2011dissociation}) enable them to survive in the harsh interstellar environment despite exposure to UV radiation and heavy ions \cite{tielens2013molecular, holm2010ions}. The ionization state of PAHs provides insights into the ionization balance of the medium, while their size and composition reveal its energetic and chemical history. 

\par The PAHs molecules are key players in interstellar chemistry, acting as catalysts that facilitate the formation of molecular hydrogen (\(H_2\)), the most abundant molecule in the interstellar medium. However, the observed abundance of \(H_2\) in cold regions cannot be explained solely by gas-phase reactions \cite{vidali2013hydrogen}. One pathway for \(H_2\) formation involves the association of hydrogen atoms on the surfaces of interstellar dust grains \cite{gould1963interstellar,wakelam2017h2}, while alternatively, PAH molecules may serve as catalysts in hydrogenation reactions \cite{boschman2012hydrogenation, klaerke2013formation, ferullo2019hydrogenated}. Furthermore, interactions between ionizing radiation and these PAHs can facilitate \(H_2\) production via dehydrogenation processes.  Joblin et al. \cite{joblin1994infrared} suggested that PAHs may serve as major catalysts in the formation of molecular hydrogen in the interstellar medium. Giard et al. \cite{giard2002photophysics} reported initial results on the photodissociation of isolated PAH cations, such as \(^{12}C_{24}H^{+}_{12}\), \(^{13}C_{1}{^{12}C_{23}}H^{+}_{12}\) and \(^{12}C_{24}H^{+}_{10}\). Recently, Champeaux et al. \cite{champeaux2014dissociation} observed that the impact of 100 keV protons on coronene molecules results in the loss of hydrogen, with one or two \(H_2\) units detected from the cation, dication and trication. They proposed that, along with the \(CH_2\) precursor, the formation of \(H_2\) is energetically more favorable than the loss of two separate hydrogen atoms.

\par The investigation of ionization cross-sections ratios, specifically, the doubly ionized to singly ionized and triply ionized to singly ionized species, in various atoms and molecules is crucial for unraveling the underlying ionization dynamics and energy deposition mechanisms during collisions \cite{manson1995ratiojob, mcguire1995ratio, dubois1988single, thompson1995single}. These ratios act as sensitive probes, revealing the efficiency of electron removal and the subsequent stabilization or fragmentation of the molecular ion. Enhanced production of multiple charged ions, as observed in studies of PAHs and other molecular systems \cite{PhysRevA.83.022704, bagdia2021ionization}, indicates that significant energy is imparted during collisions, often leading to complex fragmentation pathways. For PAHs, in particular, understanding these ratios is essential because it provides insights into their chemical resilience and reactivity under the harsh conditions of interstellar environments. This knowledge also helps in refining theoretical models-such as the continuum distorted wave–eikonal initial state (CDW-EIS) approach by incorporating molecular effects and multi-electron interactions to better describe ionization in complex systems.

\par Previously, a few PAHs have been employed in studies of ionization and fragmentation induced by photons, electrons, protons and heavy ions. Lawicki et al. \cite{PhysRevA.83.022704} studied the multiple ionization and fragmentation of coronene (\(C_{24}H_{12}\)) and pyrene (\(C_{16}H_{10}\)) molecules by the impact of \(He^{2+}\), \(O^{3+}\) and \(Xe^{20+}\) ions at low velocities (v \(\leq\) 0.6 a.u.). Bagdia et al. \cite{bagdia2021ionization} reported the ionization and fragmentation of fluorine molecules bombarded by 250 keV protons and found a substantially higher ratio of doubly to singly charged products compared to those observed for He and Ne targets. Chen et al. \cite{chen2015formation} investigated the formation of molecular ions by the impact of \(He^{+}\) ions on anthracene (\(C_{14}H_{10}\)), pyrene and coronene molecules. A signature of hydrogen loss was observed in the peaks corresponding to singly and doubly ionized PAHs. They observed a clear preference for the loss of even number of hydrogen atoms, which is important for understanding the formation of hydrogen molecules in the interstellar medium. Holm et al. \cite{holm2010ions} studied collisions of 11.25 keV \(^{3}\mathrm{He}^{+}\) and 360 keV \(^{129}\mathrm{Xe}^{20+}\) ions with weakly bound clusters of anthracene (\(C_{14}H_{10}\)) and found that these clusters tend to fragment more readily in ion collisions than other weakly bound clusters. Chen et al. \cite{chen2014absolute} also presented a scaling law for the absolute cross-sections for non-statistical fragmentation in collisions between \(PAH/PAH^{+}\) and hydrogen or helium atoms, with kinetic energies ranging from 50 eV to 10 keV.

\par In the present experiment, we investigate the interaction between low-energy protons and gas-phase coronene molecules. This study is crucial for understanding the ionization and fragmentation mechanisms of large PAHs when impacted by energetic heavy ions. Because the experimental conditions simulate proton collisions in space, the findings offer valuable insights into astrophysical processes. Detailed experimental methods and results are discussed below.

\section{Experimental details}
\begin{figure*}
\centering
\includegraphics[width=160mm]{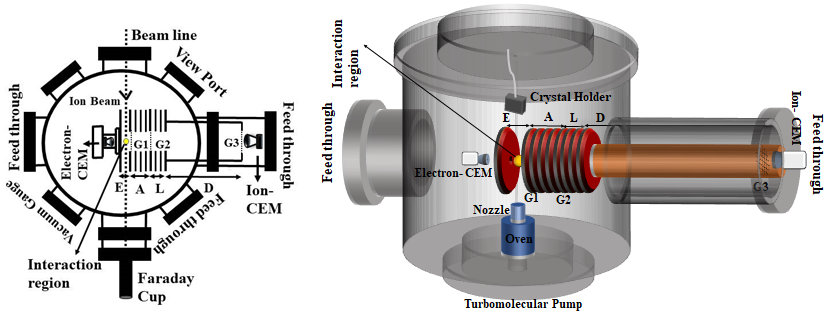}
\caption{\label{fig:schematic} Typical schematic chamber of the experimental setup (left one: upper view; right one: front view)}
\end{figure*}

\par The experiment was performed at the Electron Cyclotron Resonance Ion Accelerator (ECRIA) facility located at TIFR, Mumbai \cite{agnihotri2011ecr,mandal2019gas}. Protons with energies ranging from 100 to 300 keV were extracted from the machine to bombard on the coronene molecules. The beam was collimated using two sets of 4-jaw slits positioned about 40 cm apart, along with a 2 mm diameter collimator placed before the scattering chamber. The dimensions of the scattering chamber are 36 cm in height and 33 cm in diameter. An oven used to heat the coronene was situated at the center of the chamber, below the interaction region. A nozzle with an aspect ratio of 10 and an opening of 1 mm was employed to produce an effusive jet. The temperature of the molecules was controlled using two variacs connected in series through a step-down transformer, allowing for fine adjustments of the heating current.  The commercially available powder of the coronoene (99.99$\%$) molecule was heated up to 160 $^\circ$C. The temperature was monitored by a thermocouple attached to the oven. The vapor density of the coronene molecule could be increased by the increasing temperature. A gold crystal in a crystal holder was placed just above the interaction region to monitor the thickness of the deposited ions of coronene. This crystal was connected to the INFICON SQM-160 thickness monitor placed above the oven was used to monitor the rate of vapor flow.
\par A two-stage Wiley-McLaren type spectrometer was used with a slight modification by employing a lens \cite{biswas2021recoil} in between the acceleration region and the drift tube. This spectrometer was placed in the chamber perpendicular to the beam direction. The details of the spectrometer were published by Biswas and Tribedi \cite{biswas2021recoil}. However, a brief detail is also given here. The TOF spectrometer is divided into three regions namely, the extraction (E), acceleration (A) and drift region (D). The extraction region has two electrodes, a pusher and a puller. There are four electrodes in the acceleration region which are connected by 10 M$\Omega$ resistances. Additionally, a pair of electrodes (lens electrodes) were used in between the acceleration region and drift region to obtain better $4\pi$ collection efficiency with larger angular acceptance \cite{biswas2021recoil}. A 1.5 mm diameter hole into the pusher plate was used to collimate the generated electrons and to reduce the background. The voltage given to the pusher, puller, accelerator end, lens electrodes and drift tube were +225, -225 volts, -1905, -1515 and -1940 volts, respectively. A Ni grid (90 \% transmission) was placed at the puller, the accelerating end and at the end of the drift tube to avoid field penetration. At the other end of the spectrometer, a channel electron multiplier (CEM) (20 mm diameter) is placed to detect the recoil ions of coronene and fragmented ions. At the other end, before the pusher, another CEM of diameter 10 mm was placed to detect the ionized electrons. The electron-CEM gives the start signal and ion-CEM gives the stop signal within a selected window of 20 $\mu$s. The electron-ion coincidence spectra have been recorded using these CEMs. Typical schematic chamber of the experimental setup is shown in Fig. \ref{fig:schematic}. The vacuum in the chamber was maintained at about $1 \times 10^{-7}$ torr by using a turbo-molecular pump during the experiment. An additional pump was installed just before the chamber to enable differential pumping. This setup maintained the beamline vacuum at the order of $10^{-9}$ torr during the experiment.

\section{Continuum Distorted Wave - Eikonal Initial State (CDW-EIS) Model}

The many-electron treatment of ionization in the H$^+$ - C$_{24}$H$_{12}$ collision is reduced to a simpler one-electron description, where we consider only one electron in the molecule to be active during the collision, while the other remains bound to its initial state. Furthermore, we apply the impact parameter method, where the projectile follows a straight-line trajectory $\mathbf{R} = \boldsymbol{\rho} + \mathbf{v}t$, characterized by the constant velocity $\mathbf{v}$ and the impact parameter $\boldsymbol{\rho} \equiv (\rho, \varphi_\rho)$ \cite{mcdowell1970introduction}. The resulting single-particle scattering equation is solved within the framework of the CDW-EIS approximation introduced for atomic collisions \cite{crothers1992advances, gulyas1995cdw}). The CDW-EIS approximation belongs to the family of perturbative distorted wave methods and in order to compute the scattering observables, knowledge of multicenter initial bound and final continuum multielectronic states is required. Recently the model, based on a code developed for atomic collisions, was generalized to molecular collisions \cite{gulyas2013cdw, gulyas2016single}. The GAUSSIAN 16 quantum chemistry program package was used to determine ground-state molecular orbitals using the Linear Combination of Atomic Orbitals (LCAO) scheme and the Hartree-Fock method \cite{frisch2016gaussian}, while the positive energy (continuum) state of the molecular ion C$_{12}$H$_{24}^+$ is described on a spherically averaged potential created by the nuclei and the passive electrons. However, in the case of the C$_{24}$H$_{12}$ molecule, where 54 molecular orbitals were considered to contribute to the total yield of ionization, such a calculation is quite time-consuming. Therefore, a simpler the Complete Neglect of Differential Overlap (CNDO) method was used to describe the molecular orbitals \cite{pople1970molecular, fainstein1996impact}. The CNDO method is one of the first semi-empirical methods in quantum chemistry and it enables to evaluate the ionization probability of a given $i$th molecular orbital as sum of the constituting atomic orbitals
$$
\frac{d^2 p_i(\boldsymbol{\rho})}{d\varepsilon_{el} d\Omega_{el}} = \sum_j \sum_{\alpha_j} C_{j\alpha_j} \frac{d^2 p_{j\alpha_j}(\boldsymbol{\rho})}{d\varepsilon_{el} d\Omega_{el}}, \quad (1)
$$
where $C_{j\alpha_j}$ is the coefficients of the contributing $\alpha$th orbital of the $j$th (C or H) atom. $C_{j\alpha_j}$ is evaluated by the GAUSSIAN 16 program package \cite{frisch2016gaussian} and $\frac{d^2 p_{j\alpha_j}}{d\varepsilon_{el} d\Omega_{el}}$ is derived by the code developed for atomic ionization \cite{horbatsch1994semiclassical}.

In the framework of the independent particle model (IPM) the probabilities for the ionization of $q$ out of $N$ ($= \sum n_i$) electrons ($n_i$ is the number of electrons in the $i$th orbital) are calculated by the shell-specific binomial analysis of the single-particle probabilities (1) \cite{horbatsch1994semiclassical}. Here we consider the $q$-fold ionization probability where one of these electrons is detected at the emission energy $\varepsilon_{el}$ and the emission angle $\Omega_{el}$
$$
\frac{d^2 P_q(\boldsymbol{\rho})}{d\varepsilon_{el} d\Omega_{el}} = \sum_{q_1,\dots,q_m=0; q_1+\cdots+q_m=q}^{n_1,\dots,n_m} \prod_{i=1}^m \frac{d^2 P_{q_i}(\boldsymbol{\rho})}{d\varepsilon_{el} d\Omega_{el}}, \quad (2)
$$
\begin{align}
\frac{d^2 P_{q_i}(\boldsymbol{\rho})}{d\varepsilon_{el} d\Omega_{el}} 
&= \frac{n_i!}{q_i!(n_i - q_i)!} 
   \frac{d^2 p_i(\boldsymbol{\rho})}{d\varepsilon_{el} d\Omega_{el}} \nonumber \\
&\quad \times [p_i(\boldsymbol{\rho})]^{q_i - 1} 
   [1 - p_i(\boldsymbol{\rho})]^{n_i - q_i}. \tag{3}
\end{align}
where $m$ is the number of molecular orbitals ($m=54$ for the present case) and $p_i(\boldsymbol{\rho})$ is the total ionization probability of the $i$th orbital
$$
p_i(\boldsymbol{\rho}) = \int d\varepsilon_{el} \int d\Omega_{el} \frac{d^2 p_i(\boldsymbol{\rho})}{d\varepsilon_{el} d\Omega_{el}}. \quad (4)
$$
Like for the probability, $q$-fold ionization cross-section where the emission energy and ejection angle is observed only for one of the electrons is defined as
$$
\frac{d^2 \sigma_q}{d\varepsilon_{el} d\Omega_{el}} = \int \rho d\rho \frac{d^2 P_q(\boldsymbol{\rho})}{d\varepsilon_{el} d\Omega_{el}}, \quad (5)
$$
while the total $q$-fold ionization cross-section ($q$-TCS) is obtained as
$$
\sigma_q = \int_0^\infty d\varepsilon_{el} \int_{-1}^{+1} d(\cos \theta_{el}) \int_0^{2\pi} d\varphi_{el} \frac{d^2 \sigma_q}{d\varepsilon_{el} d\Omega_{el}}. \quad (6)
$$
Furthermore, we define the net ionization probabilities and cross-sections which are the mean values of the distributions of the corresponding $q$-fold quantities. E. g., the total cross-section (TCS) for net ionization is obtained as
$$
\sigma = \sum_{q=1}^n q\sigma_q, \quad (7)
$$
and the corresponding differential net ionization probabilities and cross-section can be defined similarly.

\begin{figure*}[hbt!]
\centering
\includegraphics[width=160mm, height=106.66 mm]{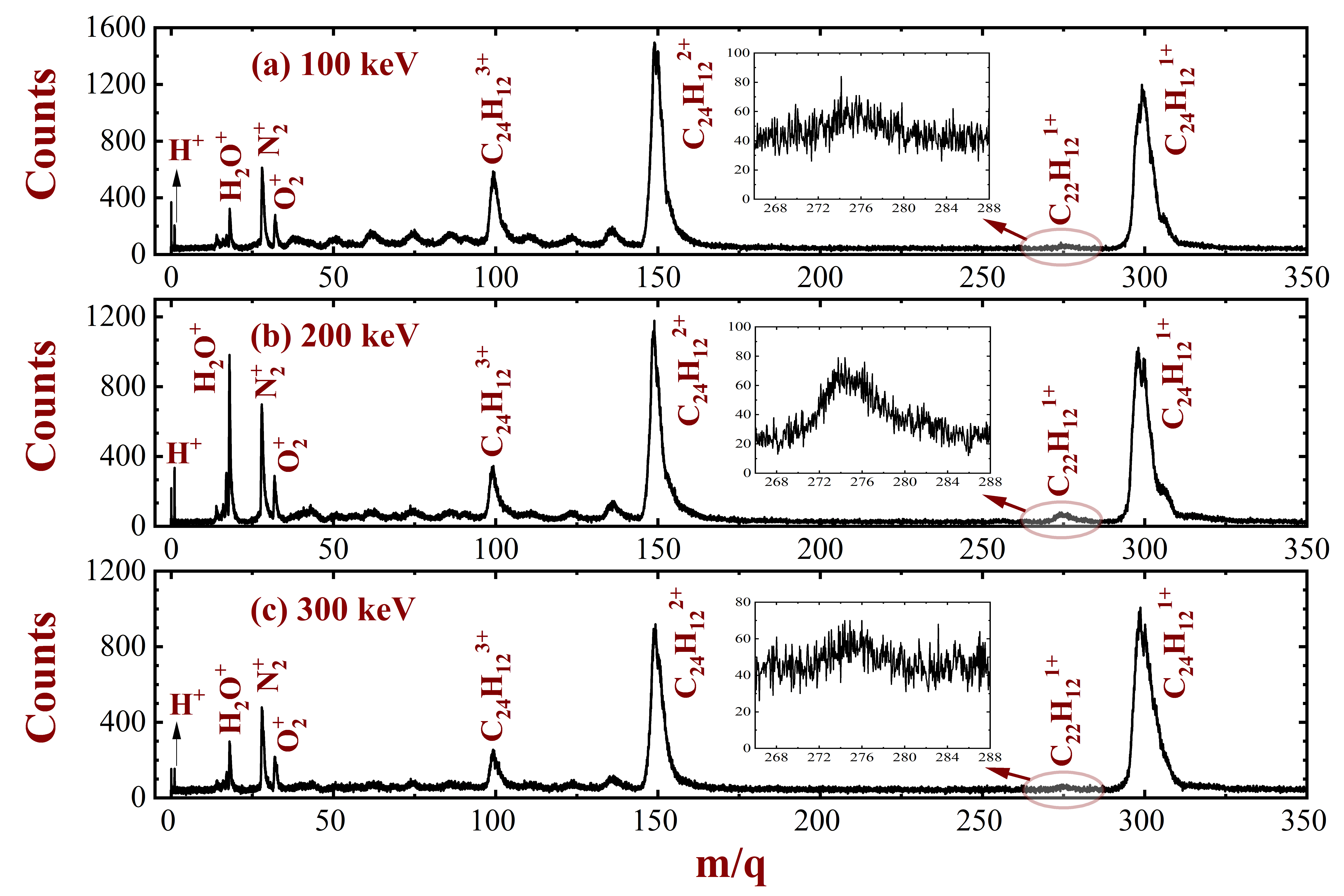}
\caption{\label{fig:spectrum}(a), (b) and (c): Typical time-of-flight (TOF) spectra of coronene as a function of mass-to-charge ratio (m/q) for proton impact at three different energies: 100 keV, 200 keV and 300 keV, respectively.}
\end{figure*}

\begin{figure*}[hbt!]
\centering
\includegraphics[width=150mm, height= 214.68mm]{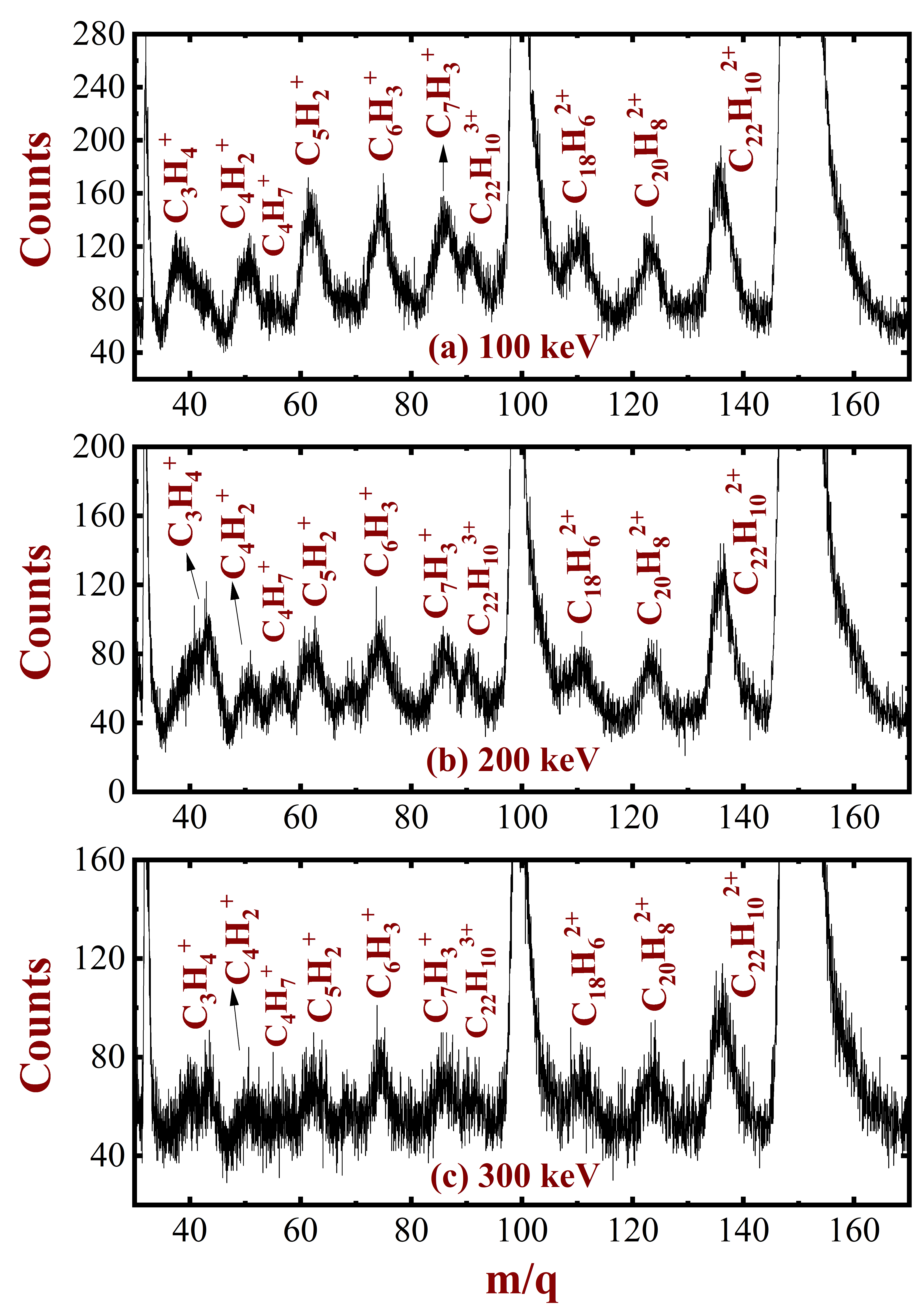}
\caption{\label{fig:zspectrum} (a), (b) and (c): Enlarged views of Fig.~\ref{fig:spectrum} (a), (b) and (c), highlighting the evaporation peaks of ${C_{24}H_{12}^{2+}}$ and fragmentation products of ${C_{24}H_{12}^{3+}}$.}
\end{figure*}

The typical mass-to-charge ratio (m/q) calibrated TOF spectra for the coronene molecule bombarded by 100, 200 and 300 keV protons are shown in Fig. \ref{fig:spectrum}. The spectra exhibit peaks corresponding to the cation ($C_{24}H_{12}^{1+}$), dication ($C_{24}H_{12}^{2+}$) and trication ($C_{24}H_{12}^{3+}$) of coronene molecules. One evaporation peak, $C_{22}H_{12}^{1+}$, corresponding to the loss of $C_2$ from the cation, is observed. The spectra also contain fragmentation products in the form of $C_nH_x$ ($n = 3 - 7$). Peaks corresponding to background gases, such as, ${O_2^{+}}$, ${N_2^{+}}$, ${H_2O^{+}}$, ${OH^{+}}$ and ${O^{+}/O_2^{2+}}$, ${N^{+}/N_2^{2+}}$, are identified. The peak of $H^{+}$ appears due to hydrogen loss from coronene molecules and $H_2O$ molecules present in the coronene powder. Fig. \ref{fig:zspectrum} presents an enlarged view of Fig. \ref{fig:spectrum}, highlighting the evaporation and fragmentation peaks of $C_{24}H_{12}^{2+}$ and $C_{24}H_{12}^{3+}$. The evaporation peaks of the parent dication, resulting from the loss of one, two and three $C_2H_2$ units ($C_{22}H_{10}^{2+}$, $C_{20}H_{8}^{2+}$ and $C_{18}H_{6}^{2+}$, respectively), are present. The evaporation peak of the trication due to the loss of a single $C_2H_2$ unit ($C_{22}H_{10}^{3+}$) is observed. 

\par In the present study, we observed coronene recoil ions up to the trication. Similarly, Lawicki et al. \cite{PhysRevA.83.022704} observed up to triply charged coronene recoil ions using $He^{2+}$ and $O^{3+}$ projectiles. However, when using highly charged $Xe^{20+}$ ions, they detected even higher charge states, including intact $C_{24}H_{12}^{5+}$ ions. The absence of such highly charged species in our experiment reflects the limited ionization capacity of protons compared to more highly charged projectiles like $Xe^{20+}$. This difference arises from the significantly higher projectile perturbation strength ($q/v$) of $\mathrm{Xe}^{20+}$ (approximately 66.7) compared to that of 100, 200 and 300~keV protons, for which $q/v$ is approximately 0.5, 0.35 and 0.29, respectively. The higher perturbation strength leads to stronger Coulomb interactions and resulting in enhanced multiple ionization. The evaporation product \( \mathrm{C}_{22}\mathrm{H}_{12}^{+} \) emerges from the cation; \( \mathrm{C}_{22}\mathrm{H}_{10}^{2+} \), \( \mathrm{C}_{20}\mathrm{H}_{8}^{2+} \) and \( \mathrm{C}_{18}\mathrm{H}_{6}^{2+} \) emerges from the dication; \( \mathrm{C}_{22}\mathrm{H}_{10}^{3+} \) emerges from the trication, all resulting from significant energy transfer to the molecule by the proton impact. The fragmentation products \(\mathrm{C}_n\mathrm{H}_x\) (with \(n = 3\text{--}7\)) originate from the trication. Since the internal energy of the trication is higher than that of the dication, it has a greater tendency to fragment into smaller products rather than evaporating neutral species. The peak identified as $C_{7}H_{3}^{+}$ can be mixed with $C_{21}H_{9}^{3+}$ i.e. $C_3H_3$ loss. The observed yield of $C_{22}H_{10}^{3+}$ ($C_2H_2$ loss) is lesser than the $C_{7}H_{3}^{+}$. So, the contribution of the $\mathrm{C_3H_3}$ loss to the $\mathrm{C_7H_3^+}$ peak should be smaller than that of $\mathrm{C_2H_2}$ loss, due to the higher binding energy of $\mathrm{C_3H_3}$ compared to $\mathrm{C_2H_2}$. Also, Dyakov et al. \cite{dyakov2006ab} studied that the $C_3H_3$ loss channel is negligible for photodissociation of azulene cation by the ab initio and Rice–Ramsperger–Kassel–Marcus (RRKM) study . So we can assume that for smaller fragmentation peaks ($C_nH_x; n= 3-7$) there is negligible contribution from $C_3H_3$ evaporation.

\subsection{Recoil-ion Production cross-sections}

\begin{figure*}[hbt!]
\centering
\includegraphics[width= 180 mm, height= 55.21mm,scale=01.0]{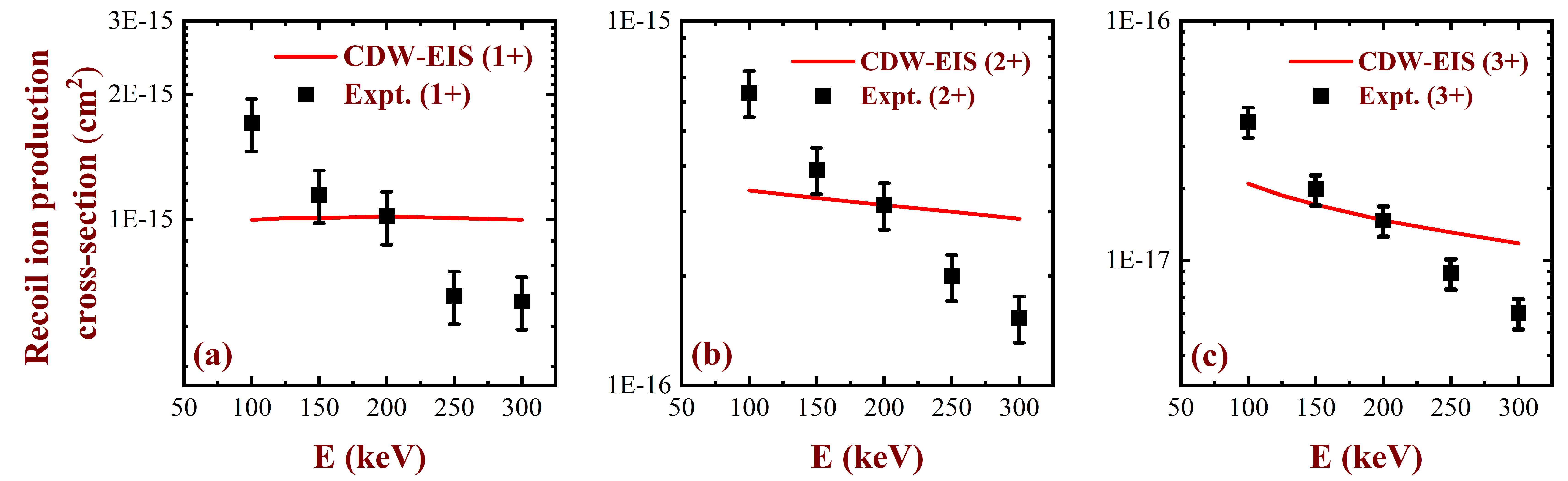}
\caption{\label{fig:yield}{Experimental (black squares) and  theoretical (CDW-EIS) ionization cross-section (red line) of the (a) singly, (b) doubly and (c) triply ionized coronene as a function of projectile energy; the experimental yield is normalized to the theoretical cross-section at 200 keV.}}
\end{figure*} 

\par The plots in Fig. \ref{fig:yield} present the production cross-sections ($cm^2$) of singly (1+), doubly (2+) and triply (3+) charged coronene recoil ions as a function of projectile energy (E in keV), comparing experimental data (black squares) with theoretical predictions from the CDW-EIS model (red lines). The uncertainties, which are within $14.5\%$, include contributions from statistical fluctuations, efficiency calibration and deposition rate estimation. The yields of 1+, 2+ and 3+ are corrected for CEM efficiency, based on the efficiency plot provided by Krems et al. \cite{krems2005channel}. The efficiency values of the channeltron for 1+, 2+ and 3+ are taken as 0.12, 0.29 and 0.44, respectively. For double and triple ionization, the probability of detecting electrons is doubled and tripled, respectively, in comparison to single ionization. Therefore, the experimental yields of 2+ and 3+ are divided by factors of 2 and 3, respectively. In the yield of the cation (1+), only single ionization (SI) is present, whereas for the dication (2+) and trication (3+), transfer ionization (TI) also contributes. The production of dications can result from either two SI events or a combination of one SI and one TI event. Similarly, the production of trications can arise from either three SI events or a combination of two SI and one TI events. In present experiment, we can not separate the direct ionization and transfer ionization. Further, the experimental yield values are normalized using the theoretical cross-sections at 200 keV to emphasize the relative energy dependence rather than absolute values.
The production cross-sections of the cation shows a decreasing trend with increasing energy and begins to saturate at 250 and 300~keV. In contrast, the theoretical prediction remains nearly flat over this energy range. The deviations observed at both lower and higher energies highlight a limitation of the model in accurately capturing the energy dependence of single ionization. For the dication and trication, the relative experimental cross-sections fall more sharply with increasing energy compared to the theoretical predictions. The enhanced relative experimental cross-sections at lower energies suggest a significant contribution from transfer ionization (TI), which is known to be more effective at lower projectile velocities. Since TI is not included in the theoretical model, this omission likely accounts for the observed deviations. At higher energies, the continued mismatch in the experimental cross-sections are further indicate absence of TI in theoretical model, as the TI cross-section itself is known to drop off more rapidly with energy, contributing to the overall sharper fall in the experimental data.

\subsection{Ratio of Recoil-ion yields: Double-to-singly charged and triple-to-singly charged ions}

\begin{figure}
\centering
\includegraphics[width=80mm, height= 114.5mm]{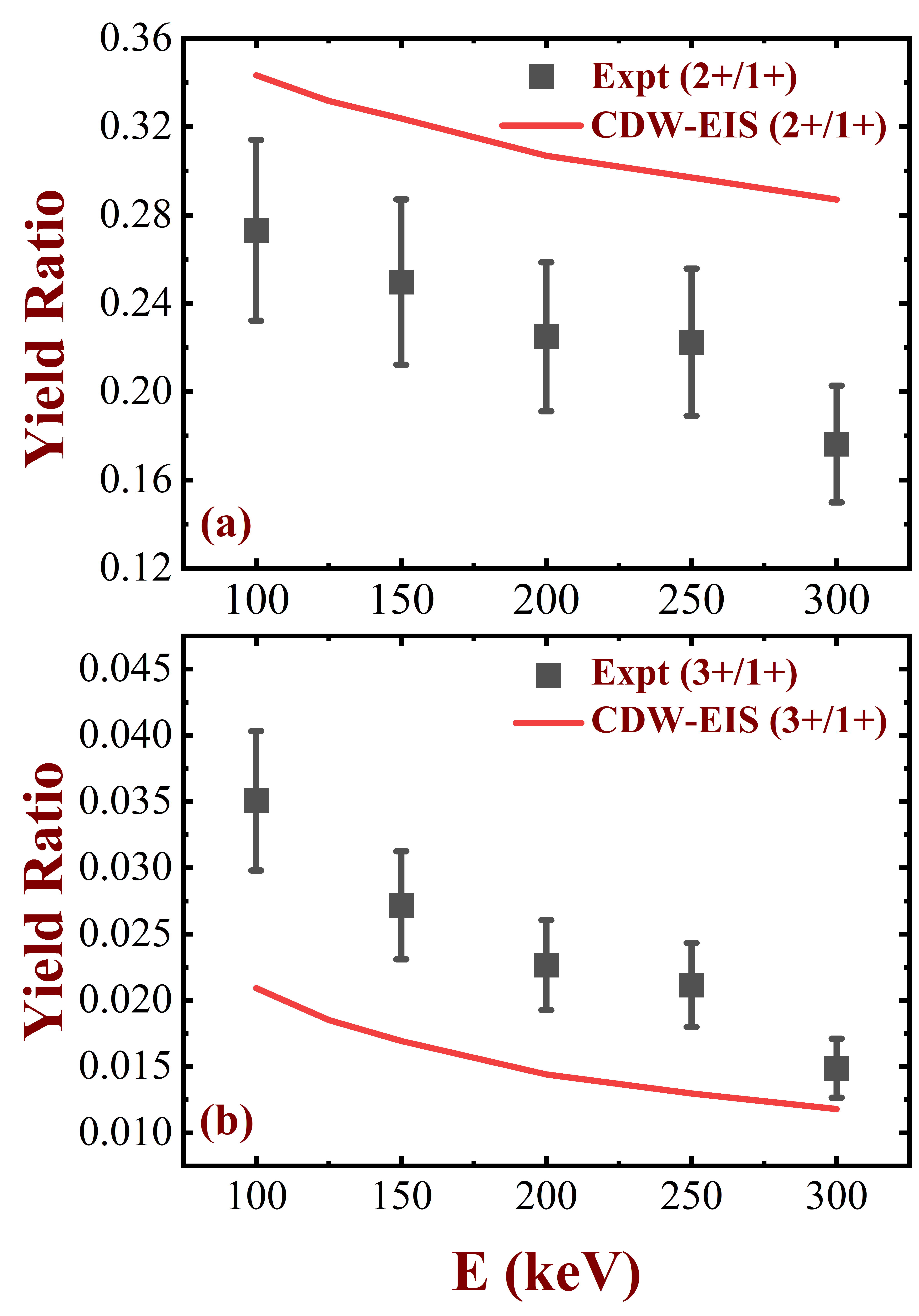}
\caption{\label{fig:ratiomerge} (a) Cross-section ratio of the doubly to singly and (b) triply to singly ionized coronene recoil ions (black squares) and the theoretical (CDW-EIS) values (red line)}
\end{figure}

\par Fig. \ref{fig:ratiomerge} presents the energy dependence of the recoil-ion cross-section ratios $R_{21} = Y(2+)/Y(1+)$ and $R_{31} = Y(3+)/Y(1+)$, (Y: yield), comparing experimental data (black squares) with theoretical predictions (red lines). The uncertainty in the data is typically $\approx 14\%$. Both $R_{21}$ and $R_{31}$ exhibit a decreasing trend with increasing projectile energy, indicating that the relative production of doubly and triply charged coronene recoil ions diminishes as the proton energy increases. The ratio of doubly and triply charged coronene relative to singly charged coronene are determined to be 0.27 and 0.035, respectively, at the lowest energy (100 keV). At the highest energy (300 keV), these ratios decrease to 0.18 and 0.015, respectively. For the $R_{21}$ (Fig. \ref{fig:ratiomerge} (a)), the theoretical values are consistently higher (on average 1.4 times) than the experimental values across all energies. This indicates that the theoretical model overestimates the probability of double ionization. 
\par In contrast, for the $R_{31}$ (Fig. \ref{fig:ratiomerge} (b)), the experimental values are higher (on average 1.4 times) than the theoretical predictions at almost all energies, particularly at lower projectile energies. This implies that the theoretical model underestimates the probability of forming triply charged coronene recoil ions. The enhanced experimental yield of $C_{24}H_{12}^{3+}$ could be attributed to additional ionization pathways, such as higher-order electron correlations or transfer ionization, which are not fully accounted for in the theoretical model. However, at the highest energy (300 keV), the experimental and theoretical values begin to converge. The sharper decline in experimental $R_{21}$ and $R_{31}$ values compared to theoretical predictions suggests that multiple ionization cross-sections decrease more rapidly with increasing energy than predicted by the model.

\par The study of double ionization of various targets by the impact of charged particles has been an area of interest for decades \cite{manson1995ratiojob,mcguire1995ratio,dubois1988single,thompson1995single,santos2005doubly,reading1989projectile,samson1990proportionality}. In earlier studies on the He atom, the double ionization (DI) process was explained through mechanisms such as two-step 1, two-step 2 and shake-off \cite{PhysRevA.83.062716, PhysRevA.57.4387, PhysRevLett.49.1153, gedeon2010, stolterfoht1997electron}. Double-to-single ionization cross-section ratios are generally low for light targets such as He, O$_2$ and N$_2$ across various ionization mechanisms. Table~\ref{tab:R21values} summarizes the measured and theoretical values of R$_{21}$ for different targets and ionization mechanisms across a broad range of energies and projectiles.
\begin{table*}[htbp]
\centering
\caption{Summary of double-to-single ionization cross-section ratios (R$_{21}$) for various targets and ionization mechanisms. R$_{21}$ values are shown as percentages only where reported as such in the literature.}
\label{tab:R21values}
\begin{tabular}{|l|l|l|c|l|}
\hline
\textbf{Target} & \textbf{Ionization Mechanism} & \makecell{\textbf{Projectile Energy/} \\ \textbf{Velocity}} & \textbf{R$_{21}$} & \textbf{Reference} \\
\hline
He & Photon impact & \makecell{2.8 keV} & (1.6 $\pm$ 0.3)\% & Levin et al. \cite{levin1991measurement} \\
O & Electron impact & \makecell{90--2000 eV} & 0.013 -- 0.024 & Thompson et al. \cite{thompson1995single} \\
O$_2$ & Electron impact & \makecell{41--400 eV} & 6.15 $\times$ 10$^{-4}$ -- 1.10 $\times$ 10$^{-2}$ & Sigaud et al. \cite{sigaud2013absolute} \\
N$_2$ & Electron impact & \makecell{200--900 eV} & 2.2 $\times$ 10$^{-2}$ -- 8.1 $\times$ 10$^{-3}$ & Sigaud et al. \cite{sigaud2018highly} \\
N$_2$ & Electron impact & \makecell{70 eV} & (7.4 $\pm$ 0.4) $\times$ 10$^{-3}$  & Shiki et al. \cite{shiki2008kinetic} \\
O$_2$ & Electron impact & \makecell{70 eV} &  (2 $\pm$ 0.4) $\times$ 10$^{-3}$ & Shiki et al. \cite{shiki2008kinetic} \\
Ar & Electron impact & \makecell{70 eV} & (48 $\pm$ 7) $\times$ 10$^{-3}$ & Shiki et al. \cite{shiki2008kinetic} \\
Ne & H$^+$ impact & \makecell{250 keV} & (13 $\pm$ 2)\% & Bagdia et al. \cite{bagdia2021ionization} \\
He & H$^+$ impact & \makecell{250 keV} & (1.1 $\pm$ 0.2)\% & Bagdia et al. \cite{bagdia2021ionization} \\
He & H$^+$ impact & \makecell{1.44 MeV/amu} & (3.3 $\pm$  0.3) $\times$ 10$^{-3}$ & Knudsen et al. \cite{knudsen1984experimental} \\
He & He$^{2+}$ impact & \makecell{0.63 MeV/amu} & (14.2 $\pm$ 1.3) $\times$ 10$^{-3}$ & Knudsen et al. \cite{knudsen1984experimental} \\
He & Li$^{3+}$ impact& \makecell{0.64 MeV/amu} & (30 $\pm$ 2.7) $\times$ 10$^{-3}$ & Knudsen et al. \cite{knudsen1984experimental} \\
He & B$^{5+}$ impact& \makecell{0.19 MeV/amu} & (237 $\pm$ 21) $\times$ 10$^{-3}$ & Knudsen et al. \cite{knudsen1984experimental} \\
He & C$^{6+}$ impact & \makecell{0.64 MeV/amu} & (98 $\pm$ 8.8) $\times$ 10$^{-3}$ & Knudsen et al. \cite{knudsen1984experimental} \\
He & O$^{8+}$ impact & \makecell{0.64  MeV/amu} & (166 $\pm$ 15) $\times$ 10$^{-3}$ & Knudsen et al. \cite{knudsen1984experimental} \\
He & He$^{2+}$ impact (CC Theory) & \makecell{0.63 MeV/amu} & 15 $\times$ 10$^{-3}$ & Barna et al. \cite{barna2005single} \\
He & Li$^{3+}$ impact (CC Theory) & \makecell{0.64 MeV/amu} & 28.5 $\times$ 10$^{-3}$ & Barna et al. \cite{barna2005single} \\
He & B$^{5+}$ impact (CC Theory) & \makecell{0.19 MeV/amu} & 204 $\times$ 10$^{-3}$& Barna et al. \cite{barna2005single} \\
He & C$^{6+}$ impact (CC Theory) & \makecell{0.64 MeV/amu} & 92 $\times$ 10$^{-3}$& Barna et al. \cite{barna2005single} \\
He & O$^{8+}$ impact (CC Theory) & \makecell{0.64 MeV/amu} & 150 $\times$ 10$^{-3}$ & Barna et al. \cite{barna2005single} \\
He & Ne$^{10+}$ impact & \makecell{$v_p/c$ $\gtrsim 0.73$} & (2.57 $\pm$ 0.10) $\times$ 10$^{-3}$ & Ullrich et al. \cite{ullrich1993high} \\
\hline
\end{tabular}
\end{table*}

\par In the present study, the higher experimental PR observed for coronene, compared to gaseous targets, may be attributed to several factors. The primary reason for the discrepancy between theoretical and experimental ratios lies in the limitations of the theoretical model. The CDW-EIS theory is based on the independent electron approximation, where one electron is considered active while others are treated as passive. Notably, this model does not account for multi-electron correlation effects and includes only direct ionization (DI), while transfer ionization (TI) is not considered. The unexpectedly large yields of doubly and triply ionized recoil ions from PAH molecules can be partly attributed to the highly correlated electron cloud and the possible excitation of plasmonic states during collisions. Such plasmonic behavior has been reported for coronene and fluorene in heavy-ion collisions \cite{biswas2015plasmon, biswas2017differential}. This plasmonic behavior, known as giant dipole plasmon resonance (GDPR), arises from the collective oscillations of delocalized \(\pi\) electrons in the extensive conjugated systems of PAHs. The GDPR enables PAHs to absorb and redistribute energy more efficiently during ionizing collisions, thereby enhancing multiple electron ejection. In recent experiments on fullerene, the influence of strong electron electron correlation was responsible for GDPR \cite{kasthurirangan2022observation}. In contrast, smaller atomic and diatomic systems, which possess more localized electronic structures, do not support these collective excitations, leading to significantly lower multi-ionization yields. 
\subsection{Evaporation and fragmentation yield}
\begin{figure}
\centering
\includegraphics[width=80mm, height= 61.26mm]{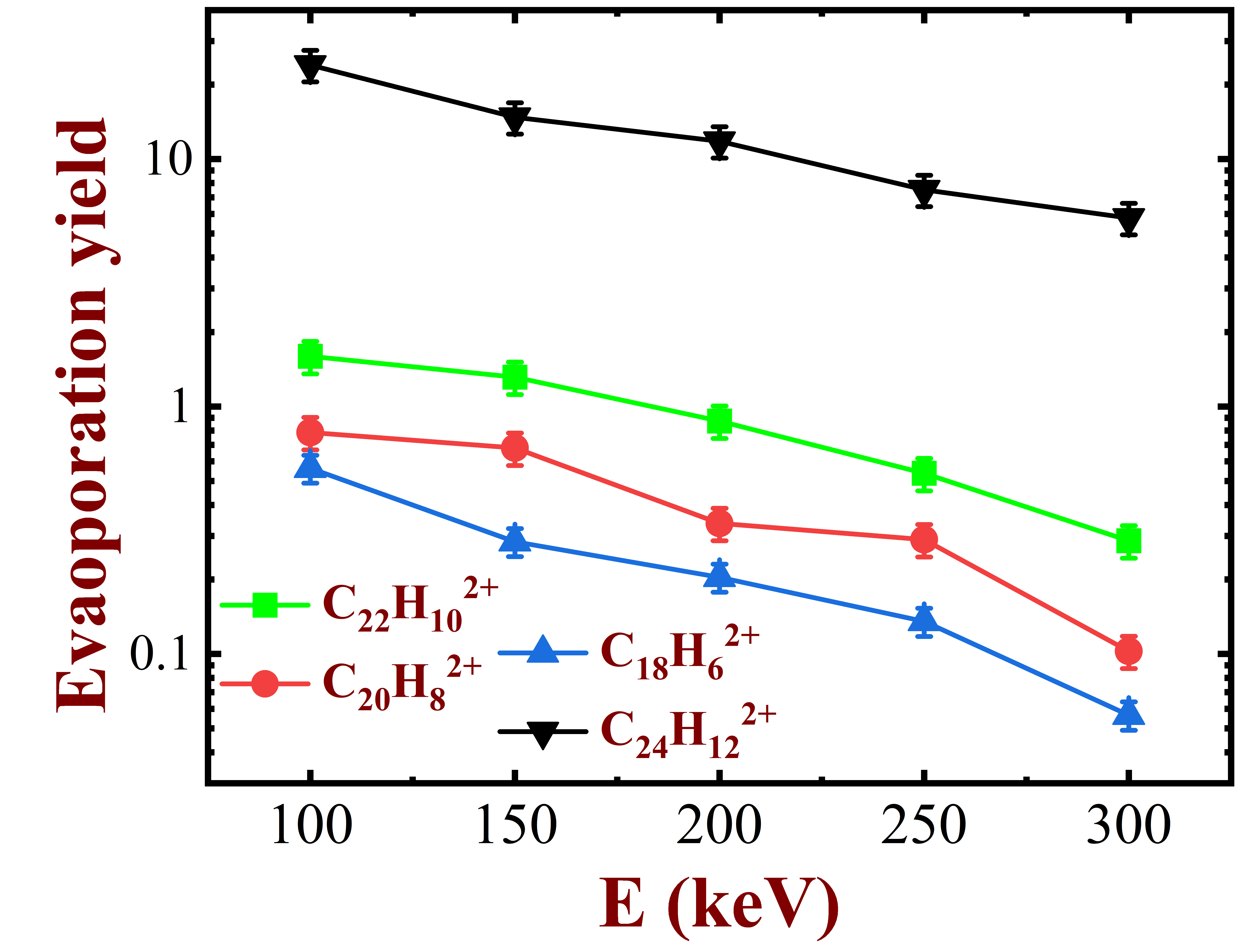}
\caption{\label{fig:evapyield} Evaporation yields of evaporation products as a function of projectile energy. 
$\mathrm{C_{22}H_{10}^{2+}}$ (green squares), $\mathrm{C_{20}H_{8}^{2+}}$ (red circles), $\mathrm{C_{18}H_{6}^{2+}}$ (blue upward triangles) and $\mathrm{C_{24}H_{12}^{2+}}$ (black downward triangles). Lines are drawn to guide the eye.}
\end{figure}
Fig.~\ref{fig:evapyield} shows the evaporation yields of three evaporation products $\mathrm{C_{22}H_{10}^{2+}}$, $\mathrm{C_{20}H_{8}^{2+}}$ and $\mathrm{C_{18}H_{6}^{2+}}$ as a function of projectile energy. The yield of the intact parent dication is also plotted for comparison. All product yields are corrected to their respective detection efficiencies. The maximum percentage uncertainty in the yield is 15\%. The relative yield attenuation (percentage decrease from 100~keV to 300~keV) for the evaporation fragments C$_{22}$H$_{10}^{2+}$, C$_{20}$H$_{8}^{2+}$ and C$_{18}$H$_{6}^{2+}$ is found to be 82\%, 87\% and 90\%, respectively. The parent dication yield shows a comparatively smaller attenuation of 76\% over the same energy range, indicating that the energy-dependent yield attenuation is more pronounced for the evaporated fragments than for the parent ion. At higher projectile energies, the interaction time between the ion and the molecule is shorter, which causes less energy transfer. Since evaporation requires the accumulation of sufficient internal energy within the molecule, the yields of these evaporated fragments decrease more rapidly at higher energies. On the other hand, the yield of the intact dication falls more slowly because it is not mainly driven by internal energy thresholds.

\begin{figure*}
\includegraphics[width= 180 mm, height= 55.21mm,scale=01.0]{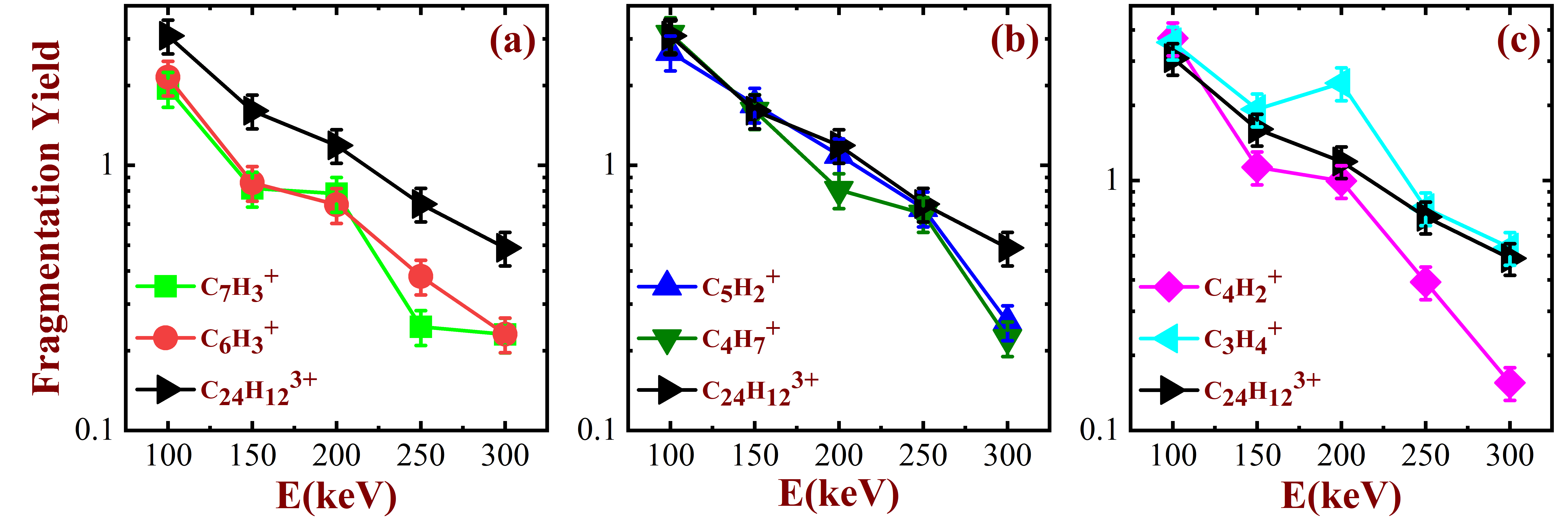}
\caption{\label{fig:fragyield} Fragmentation yields of various fragment ions as a function of projectile energy. (a) $\mathrm{C_{7}H_{3}^{+}}$ (light green squares), $\mathrm{C_{6}H_{3}^{+}}$ (red circles); (b) $\mathrm{C_{5}H_{2}^{+}}$ (blue upward triangles),  $\mathrm{C_{4}H_{7}^{+}}$ (green downward triangles); (c) $\mathrm{C_{4}H_{2}^{+}}$ (magenta diamonds), $\mathrm{C_{3}H_{4}^{+}}$ (cyan leftward triangles); $\mathrm{C_{24}H_{12}^{3+}}$ (black rightward triangles). Lines are drawn to guide the eye.}
\end{figure*}

\par Fig.~\ref{fig:fragyield} shows the fragmentation yields of the fragment ions $\mathrm{C_7H_3^+}$, $\mathrm{C_6H_3^+}$, $\mathrm{C_5H_2^+}$, $\mathrm{C_4H_7^+}$, $\mathrm{C_4H_2^+}$ and $\mathrm{C_3H_4^+}$ as a function of projectile energy. The yield of the intact parent trication is also plotted for comparison. All product yields are corrected for their respective detection efficiencies. The maximum percentage uncertainty in the yield is 15\%. The percentage decrease in yield from 100 to 300~keV for the fragment ions is as follows: $\mathrm{C_7H_3^+}$ (88\%), $\mathrm{C_6H_3^+}$ (89\%), $\mathrm{C_5H_2^+}$ (90\%), $\mathrm{C_4H_7^+}$ (93\%), $\mathrm{C_4H_2^+}$ (96\%) and $\mathrm{C_3H_4^+}$ (85\%). For the parent ion $\mathrm{C_{24}H_{12}^{3+}}$, the percentage decrease is 84\%. The changes in fragmentation yield with increasing projectile energy show that different fragments respond differently when the molecule is struck by fast-moving projectiles. Some fragment ions, such as $\mathrm{C_4H_2^+}$, $\mathrm{C_4H_7^+}$, $\mathrm{C_5H_2^+}$, $\mathrm{C_6H_3^+}$ and $\mathrm{C_7H_3^+}$, show a much sharper decrease in yield compared to the parent ion. On the other hand, fragments like $\mathrm{C_3H_4^+}$ show a decrease in yield that is quite similar to the parent ion and this ion also shows a noticeable peak in yield around 200~keV. In general, the degree to which each fragment's yield decreases depends on how much energy is required to produce it, fragments that require higher internal excitation drop off faster, while those formed via more stable or low-energy pathways persist longer.

\subsection{Hydrogen loss}

\begin{figure*}
\centering
\includegraphics[width=160mm]{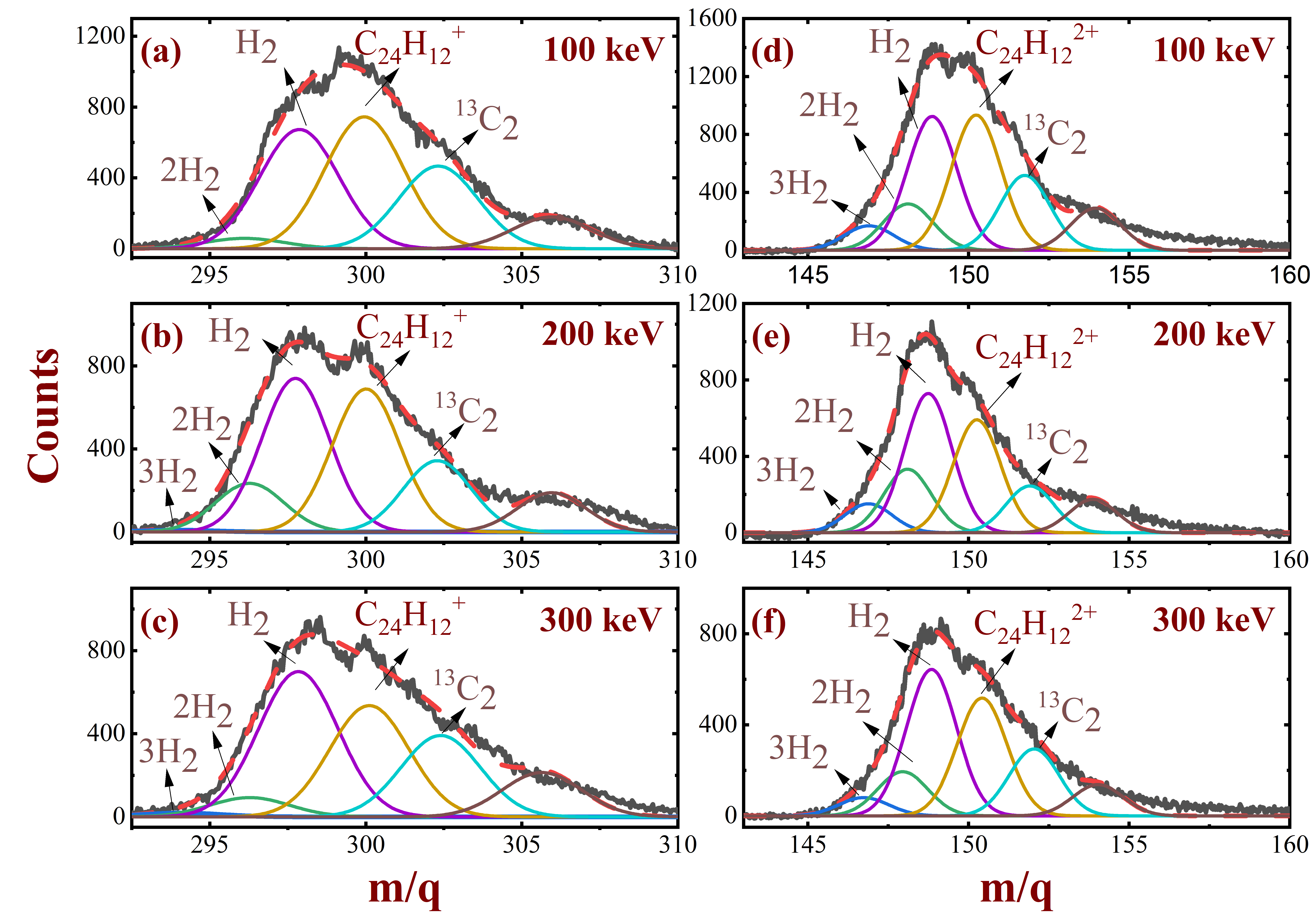}
\caption{\label{fig:fittedpeaks} Singly and doubly ionized coronene showing $nH_2$ loss components by multi-Gaussian fitting; black solid curve: observed peak structure, red dashed curve: fitted peak structure, (a), (b) and (c): cation peak structure, (d), (e) and (f): dication peak structure}
\end{figure*}

\begin{table*}
\centering
\caption{\label{tab:Hloss} Contribution (in percentage) of the parent ion, n$H_2$ losses (n = 1 to 3) and carbon isotopes in the singly and doubly ionized coronene recoil ions at 100, 200 and 300 keV projectile energies, relative to the total yield}
\begin{tabular}{l l l l l l l l l l l l l l l l l l l l l l l l}
\hline
\makecell{E (keV)} & \makecell{species/ion}& \makecell{parent ion} & \makecell{1 $H_2$} & \makecell{2 $H_2$} &  \makecell{3 $H_2$}   \\  

\makecell{100 keV} &\makecell{$C_{24}H_{12}^{1+}$}& \makecell{38} & \makecell{35} & \makecell{3}& \makecell{--} \\

\makecell{}&\makecell{$C_{24}H_{12}^{2+}$}& \makecell{33} & \makecell{32}& \makecell{11}& \makecell{6}\\
  
\hline
\makecell{200 keV} &\makecell{$C_{24}H_{12}^{1+}$}& \makecell{34} & \makecell{37}& \makecell{12}& \makecell{0.5} \\

\makecell{}&\makecell{$C_{24}H_{12}^{2+}$}& \makecell{29} & \makecell{36}& \makecell{16}& \makecell{7}\\

\hline 
\makecell{300 keV}& \makecell{$C_{24}H_{12}^{1+}$}& \makecell{31} &\makecell{40}& \makecell{5}& \makecell{1} \\

\makecell{}&\makecell{$C_{24}H_{12}^{2+}$}& \makecell{30} & \makecell{37}& \makecell{11}& \makecell{5}\\

\hline
\end{tabular}
\end{table*}

\par The enlarged peak structures around coronene cation, dication and trication for 100, 200 and 300 keV projectiles are shown in Fig. \ref{fig:fittedpeaks}. The peak structures have been fitted using a multi-Gaussian function, with a maximum fitting uncertainty of $m/q = \pm 0.4$. The peaks to the left of the cation peaks (m/q = 300) correspond to the loss of an even number of hydrogen molecules ($H_2$) (1–3 $H_2$ losses). At 200 and 300 keV, the contribution 3 $H_2$ loss channel is so low that it can not be seen clearly. Similarly, the peak structure around the dication peaks (m/q = 150) exhibits multiple peaks, where the left-side peaks correspond to sequential hydrogen losses (up to 3 $H_2$). The peaks to the right of the m/q = 300 and 150 positions can be partially attributed to isotopic contributions from $^{13}$C. The peak observed at the extreme right remains unidentified and its origin currently unknown. The 1H$_2$ loss peak is distinctly visible, peaks corresponding to 2H$_2$ and 3H$_2$ losses are not as clearly resolved. Incorporating the 2H$_2$ and 3H$_2$ components reduces chi-square ($\chi^2_{red}$) values and improves the overall fit quality. Fits performed without them lead to significantly higher $\chi^2_{red}$ values and show systematic residuals, particularly in the lower mass tail of the main peak. Table~\ref{tab:Hloss} presents the contributions of the parent ion, nH$_2$ losses (where n=1 to 3) and carbon isotopic peaks to the total yield of singly and doubly ionized coronene recoil ions.  At 100 keV, for peak structure around cation, approximately 38\% of the total yield corresponds to the $C_{24}H_{12}^+$, while significant fractions of the total signal are attributed to the loss of one ($35\%$) and two ($3\%$) $H_2$ loss channel. For peak structure around dication, 33\% of the total yield corresponds to the $C_{24}H_{12}^{2+}$, 32\% is associated with the loss of one H$_2$ molecule, 11\% with the loss of two and 6\% with the loss of three H$_2$ molecules. The observation of significantly higher three H$_2$ loss channel for the dication indicates a higher internal energy compared to the singly charged cation. With increasing projectile energy, the contribution of intact parent ions decreases for both singly and doubly charged coronene and the contribution from H$_2$ loss fragments increases (shown in Table~\ref{tab:Hloss}). The contribution from total H$_2$ loss fragments for both cation and dication is found to peak at 200 keV and then decrease at 300 keV. The increased H$_2$ loss at 200 keV suggests that the energy transfer at this value is particularly effective for breaking C–H bonds and indicates an optimal energy for this dissociation pathway. 

\par In our TOF mass spectra, we observed prominent fragmentation channels involving the loss of $C_2H_2$ and $H_2$, while H-loss was notably absent. Lawicki et al. Lawicki et al. \cite{PhysRevA.83.022704} reported no H loss in the cation and up to 4 H losses (only even) in the dication of coronene following impact with $Xe^{20+}$ ions. Furthermore, they observed up to 4 H losses (both even and odd) in the cation and up to 6 H losses (only even) in the dication of coronene when impacted by $He^{2+}$ ions. In previous studies \cite{zhen2015laboratory,zhen2016vuv,joblin2020photo}, various PAH cations were exposed to VUV radiation, resulting in H loss from the cations. Hydrogen dissociation from polycyclic aromatic hydrocarbons (PAHs) has been theoretically studied, where H loss is supported for naphthalene cations \cite{jolibois2005}. Champeaux et al. \cite{champeaux2014dissociation} observed only even-numbered H losses (up to 4 $H$ losses) in coronene cation, dication and trication upon impact with 100 keV protons. Since only even-numbered H losses were present, they suggested that the elimination of $H_2$ is energetically more favorable than the loss of two individual hydrogen atoms along with a $CH_2$ precursor. Paris et al. ~\cite{paris2014} have shown using DFT calculations that among various hydrogen loss pathways in coronene, neutral H$_2$ loss is the preferential channel for neutral, singly, doubly and triply-charged species. Our experimental observation of only H$_2$ emission is therefore in good agreement with their prediction. In theoretical study reported by Simon et al. \cite{simon2017dissociation}, at 40~eV internal energy, the dissociation of C$_{24}$H$_{12}^+$ shows that H atom loss is the dominant initial fragmentation channel, with C$_{24}$H$_{11}^+$ (m = 24, n = 11) reaching a peak abundance of $\sim$20\% at around 50~ps, but then gradually decreasing to less than $\sim$10\% by 200~ps. In contrast, H$_2$ loss, corresponding to the formation of C$_{24}$H$_{10}^+$ (m = 24, n = 10), increases more gradually but steadily, overtaking H loss around $\sim$200~ps and reaching $\sim$10\% at 200~ps. This shift indicates that while H loss dominates early, H$_2$ loss becomes the more significant pathway at later times. At 30~eV internal energy, the dissociation of C$_{24}$H$_{12}^+$ is significantly slower and more limited compared to 40~eV. H atom loss, producing C$_{24}$H$_{11}^+$ (m = 24, n = 11), is found, reaching an abundance of approximately 9\% after 500~ps. There is no clear evidence of H$_2$ loss at this energy. To fully understand the dissociation dynamics of the coronene molecule under energetic processing, it is essential to investigate its behavior at higher internal energies. This will be particularly relevant for our case, where the energy imparted to coronene by proton impact at 100~keV has been estimated to be 77~eV, as reported by Champeaux et al. \cite{champeaux2014dissociation}. Such energy levels significntly exceed the 40~eV and 30~eV studied in Simon's work. Their results indicate that at 40~eV, H$_2$ loss becomes the dominant fragmentation pathway at longer dissociation times, overtaking H atom loss. Therefore, at internal energies of 77~eV and above, H$_2$ production is expected to dominate even more prominently. Our results are consistent with previous theoretical and experimental studies, which indicate that H$_2$ loss is the preferred dissociation pathway in coronene. The presence of distinct peaks corresponding to H$_2$ loss events in our spectra further supports this mechanism. At 30~eV, Simon et al.\cite{simon2017dissociation} also observed negligible C$_2$H$_2$ loss. Although their study reported that C$_2$H$_2$ loss becomes the dominant fragmentation channel for coronene cations at 40~eV internal energy.

\section{Conclusions}
The time of flight spectrum has been recorded for the coronene molecule by the bombardment of protons using a two-stage Wiley-Mclaren type spectrometer. We detected up to trication of coronene along with evaporation and fragmentation peaks. The energy dependence of the relative cross-section of cation, dication and trication show the sharper decline with increasing energy than the theoretical model (CDW-EIS). The \( \mathrm{2{+}/1{+}} \) cross-section ratio is overestimated by the theory by a factor of approximately 1.4, while the \( \mathrm{3{+}/1{+}} \) ratio is underestimated by a similar factor. Dramatically enhanced $2{+}/1{+}$ and $3{+}/1{+}$  ratios are found which are much larger than those for typical gaseous targets which is partly be due to highly correlated electron cloud and the plasmonic excitation (studied earlier). The evaporation products, such as, the C$_{2}$H$_{2}$-loss peaks, show strong energy dependence, with yields decreasing by an average of 85\% as the projectile energy increases from 100 to 300~keV. Also, fragmentation products (C$_n$H$_x^{+}$ ions) show a steep decline in yield, indicating a reduced internal energy accumulation at higher velocities. In the cation and dication spectra, the loss of multiple hydrogen atoms (up to six) is clearly observed. Only even-numbered hydrogen losses are observed in the present study, consistent with previously reported experimental and theoretical results which strongly support the preferential elimination of $H_2$ rather than individual hydrogen atoms. The loss of the hydrogen in PAHs through ionizing radiation may be a possible source of the presence of the $H_2$ in space. These findings not only highlight the unique ionization and fragmentation dynamics of large PAH molecules but also reinforce their potential role as active catalysts in interstellar molecular processes. The experimentally observed correlation between ionization state, hydrogen loss and molecular dissociation processes provides valuable benchmarks for theoretical models. Future studies involving different PAHs and projectile species can further unravel the intricate pathways of ion-induced chemistry in astrochemical environments.

\section{Acknowledgment}
The authors are thankful to the staff of the ECRIA-laboratory of TIFR, particularly Nilesh Mhatre, D. Pathare and S. Manjrekar.



\nocite{*}
\bibliographystyle{apsrev4-2}
\bibliography{apssamp}
\end{document}